\documentclass[10pt,twoside]{epnt1p}
\usepackage{epsfig}

\usepackage{amssymb}
\usepackage{amsmath}

\usepackage{url}
\begin{document}

\begin{frontmatter}


\title{Ultra High Energy Neutrino Signature in Top-Down Scenario}


\author[address1]{R. Aloisio},

\address[address1]{INFN - Laboratori Nazionali del Gran Sasso, I-67010
  Assergi (AQ), Italy }

\begin{abstract}
Neutrinos are the best candidates to test the extreme Universe and ideas
beyond the Standard Model of particle Physics. Once produced, neutrinos do 
not suffer any kind of attenuation by intervening radiation fields like the 
Cosmic Microwave Background and are not affected by magnetic fields. In this 
sense neutrinos are useful messengers from the far and young Universe. In the
present paper we will discuss a particular class of sources of Ultra High 
Energy Cosmic Rays introduced to explain the possible excess of events with 
energy larger than the Graisen-Zatsepin-Kuzmin cut-off. These sources, 
collectively called top-down, share a common feature: UHE particles are
produced in the decay or annihilation of superheavy, exotic, particles. 
As we will review in the present paper, the largest fraction of Ultra 
High Energy particles produced in the top-down scenario are neutrinos.
The study of these radiation offers us a unique opportunity to test the 
exotic mechanisms of the top-down scenario.
\end{abstract}



\end{frontmatter}

\section{\label{sec_in} Introduction}

The Ultra High Energy (UHE) neutrino detection is one the most important 
step forward in the Cosmic Ray (CR) Physics. The discovery of neutrinos with 
energy larger than $10^{17}$ eV will start the neutrino astronomy, enabling 
the observation of the most distant and powerful sources in the Universe. 
The existence of neutrinos with such high energy is intimately related to 
the observation of Ultra High Energy Cosmic Rays (UHECR). 

Soon after the discovery of the Cosmic Microwave Background (CMB) radiation 
it was shown that the flux of UHECR, with an energy larger than $10^{18}$ eV, 
should be characterized by a sharp steepening at energies $\sim 10^{20}$~eV, 
due to the absorption processes on the CMB radiation. This effect 
is the well known Graisen-Zatsepin-Kuzmin (GZK) cut-off \cite{GZK}. 
After a few decades of observations the detection of the GZK steepening 
is still one of the major open problems in UHECR physics and the experimental 
data are not conclusive. The 11 Akeno Grand Air Shower Array (AGASA) events 
with energy larger than $10^{20}$ eV \cite{expCR} contradict the expected 
suppression of the UHECR spectrum. On the other hand the HiRes data seem to 
be consistent with the GZK cut-off picture \cite{expCR}. If the UHECR 
primaries are protons and if they propagate rectilinearly, as the claimed 
correlation with BL-Lacs at energy $4-8\times 10^{19}$ eV implies \cite{TT}, 
than their sources must be seen in the direction of the highest energies 
events with energies up to $2-3 \times 10^{20}$ eV detected by HiRes, 
Fly's Eye and AGASA \cite{expCR}. At these energies the proton attenuation 
length is only about $20-30$ Mpc and no counterparts in any frequency band 
was observed in the direction of these UHECR events. This is a strong 
indication that CR particles with energies larger than $10^{20}$ eV may 
have a different origin from those with lower energies. 

Models of origin of UHECR fall into two categories, top-down and 
bottom-up. In the bottom-up scenario UHECR originates from cosmic 
accelerators. In these accelerators particles of relatively low energy are 
brought to UHE through multiple interactions at the source. The most 
promising accelerators known in nature are based on the diffusive shock 
acceleration mechanism, in which the particle acceleration is realized through 
multiple interactions with a shock front. This mechanism works fairly well in 
Super Nova Remnants (SNR) that are believed to be the responsible for the 
acceleration of Galactic Cosmic Rays of energy $E<10^{18}$ eV (for a review 
see \cite{Hillas}). At the highest energies different bottom-up scenario have 
been proposed, among them, the most promising, are those in which acceleration 
is realized through the interaction with a relativistic shock front in Active 
Galactic Nuclei (AGN) and Gamma Ray Bursts (GRB). In the framework of 
bottom-up 
models the observed UHECR flux should show the predicted GZK steepening and 
UHE neutrinos are produced by the interaction of UHECR with different 
backgrounds, at the source and during their journey to us. These are the 
so-called cosmogenic neutrinos (first proposed in \cite{BereZats}) that we 
will not discuss here.

The presence of an excess of events as claimed by AGASA inspired the 
introduction of several exotic models for the production of UHECR. These 
models, collectively called top-down, explain the excess of AGASA and give 
also a clear explanation for the lacking of any counterpart of the highest 
energy events. Many different ideas have been proposed among top-down models:
strongly interacting neutrinos \cite{nu} and new light hadrons \cite{gluino} 
as unabsorbed signal carriers, $Z$-bursts \cite{Z}, Lorentz-invariance 
violation \cite{Lorentz}, Topological Defects (TD) (see \cite{TD} for a 
review), and Superheavy Dark Matter (SHDM) (see \cite{SHDM} for a review).

In the present paper we will concentrate our attention on the two last models 
that show common features: UHE particles are produced in the decay of 
superheavy particles, that we shall call collectively $X$ particles,
with a typical mass of the order of the Grand Unified energy scale 
$M_{GUT}\simeq 10^{24}$ eV. In the case of TD the $X$ particle once
produced, by the internal dynamics of the defect or through the interaction
of different defects, immediately decays. While in the case of SHDM the $X$
particle itself is long-lived contributing to the Dark Matter of the universe. 

From the point of view of elementary particle physics the $X$ particle 
decay process proceed in a way similar to $e^+e^-$ annihilation into hadrons:
two or more off-mass-shell quarks and gluons are produced and they initiate 
a QCD cascade. Finally the partons are hadronized at the confinement radius. 
Most of the hadrons in the final state are pions and thus the typical
prediction of all these models is the dominance of neutrinos and photons at the
highest energies $E \ge 5\times 10^{19}$~eV. It is important to stress here
that these models predict neutrino fluxes most likely within reach of the 
first generation neutrino telescopes such as AMANDA, and certainly detectable 
by future kilometer-scale neutrino observatories \cite{GHT}.

\section{\label{sec_1} Hadrons spectrum in X decay}

The first step to determine the neutrino flux produced in the 
decay of $X$ particles is the determination of the hadron 
spectrum. Moreover, this evaluation is particularly important because 
it represents a direct signature of the production mechanism that, in 
principle, can be detected experimentally. As discussed in the introduction, 
the mass of the decaying particle, $M_X$, that represents the total CMS 
energy $\sqrt{s}$, is in the range $10^{13}$ -- $10^{16}$~GeV. 

The existing QCD Monte Carlo (MC) codes become numerically unstable at much 
smaller energies, e.g., at $\sqrt{s} \sim 10^7$~GeV and the computing time 
increases rapidly going to larger energies. In this section we will briefly 
review the main results obtained, in the computation of the top-down spectrum 
of UHE particles, using two different computational techniques: one based on a 
new MC scheme \cite{ABK,BereKach} and the other based on the 
Dokshitzer-Gribov-Lipatov-Altarelli-Parisi (DGLAP) evolution equations 
\cite{ABK,previous}. In both cases SUSY is included in the computation.

Monte Carlo simulations are the most physical approach for high
energy calculations which allow to incorporate many important physical
features as the presence of SUSY partons in the cascade and
coherent branching. The perturbative part of a QCD Monte Carlo simulation 
is quite standard with the inclusion of SUSY. For the non-perturbative 
hadronization part an original phenomenological approach is used in 
Ref.~\cite{ABK}. The fragmentation of a parton $i$ into an hadron $h$ is 
expressed through perturbative fragmentation function of partons 
$D_i^j(x,M_X)$, that represents the probability of fragmentation of a parton 
$i$ into a parton $j$ with momentum fraction $x=2p/M_X$, convoluted with the 
hadronization functions $f_j^h(x,Q_0)$ at scale $Q_0$, that is understood 
as the fragmentation function of the parton $i$ into the hadron $h$ at the 
hadronization scale $Q_0\simeq 1.4$ GeV \cite{ABK}. To obtain the 
fragmentation functions of hadrons one has:

\begin{equation}
D_i^h(x,M_X)=
\sum_{j=q,g}\int_x^1\frac{dz}{z}D_i^j(\frac{x}{z},M_X)f_j^h(z,Q_0)
\label{hfunc}
\end{equation}

where the hadronization functions do not depend on the scale
$M_X$. This important property of hadronization functions allows the  
determination of $f_i^h(x,Q_0)$ from the available LEP data, $D_i^h(x,M_X)$ 
at the scale $M_X=M_Z$. 

The fragmentation functions $D_i^h(x,M_X)$ at a high scale $M_X$ can
be calculated also evolving them from a low scale, e.g. $M_X=M_Z$, where they 
are known experimentally or with great accuracy using the MC scheme. This 
evolution is described by the Dokshitzer-Gribov-Lipatov-Altarelli-Parisi 
(DGLAP) equation \cite{DGLAP} which can be written as

\begin{equation}
\partial_t D_i^h=\sum_j\frac{\alpha_s(t)}{2\pi}P_{ij}(z)\otimes
D_j^h(x/z,t)\,,
\label{DGLAP-eq}
\end{equation}

where $t=\ln(s/s_0)$ is the scale, $\otimes$ denotes the convolution
$f\otimes g=\int_z^1 dx/x f(x)g(x/z)$, and $P_{ij}$ is the splitting function 
which describes the emission of parton $j$ by parton $i$. Apart from the 
experimentally rather well determined quark fragmentation function 
$D_q^h(x,M_Z)$, also the gluon fragmentation function $D_g^h(x,M_Z)$ 
is needed for the evolution of Eq.~(\ref{DGLAP-eq}). The gluon FF can be taken 
either from MC simulations or from fits to experimental data, in particular
to the longitudinal polarized $e^+e^-$ annihilation cross-section and
three-jet events. The first application of the DGLAP method for the 
calculation of hadron spectra from decaying superheavy particles has been 
made in Refs.~\cite{previous}. The most detailed calculations have been 
performed by Barbot and Drees \cite{previous}, where more than 30 different 
particles were allowed to be cascading and the mass spectrum of the SUSY 
particles was taken into account. The results obtained with the two different 
techniques discussed above agree fairly well \cite{ABK}. The accuracy in the 
hadron spectrum calculations has reached such a level that one can consider 
the spectral shape as a signature of the model. The predicted hadron spectrum 
is approximately $\propto dE/E^{1.9}$ in the region of $x$ relevant for UHECR 
observations.

\section{\label{sec_2} Spectra of Neutrinos, Photons and Nucleons}

The spectra of neutrinos and photons produced by the decay of
superheavy particles are of practical interest in high energy
astrophysics and can be computed from the decay of charged pions
\cite{ABK}. The FFs for charged pions and protons+antiprotons 
can be determined, following \cite{ABK}, from the FFs of hadrons 
$D_h$ simply introducing the ratios $\varepsilon_N(x)$ and 
$\varepsilon_{\pi}(x)$ as: $D_N(x) = \varepsilon_N(x) D_h(x)$ and 
$ D_{\pi}(x) = \varepsilon_{\pi}(x) D_h(x)$. The spectra of pions and 
nucleons at large $M_X$ have approximately the same shape as the hadron 
spectra, and one can use in this case $\varepsilon_{\pi}=0.73\pm 0.03$ and 
$\varepsilon_{N}=0.12\pm 0.02$ \cite{ABK}.   

An interesting feature of the up-dated calculations performed in 
\cite{ABK} and by Barbot and Drees in \cite{previous} is the ratio of 
photons to nucleons, $\gamma/N$. At $x\sim 1\times 10^{-3}$ this ratio 
is characterized by a value of 2 -- 3 only \cite{ABK}. This result has an 
important impact for SHDM and topological defect models because the fraction 
of nucleons in the primary radiation increases. However, in both models 
photons dominate (i.e. their fraction becomes $\ge 50\%$) at 
$E\ge (7-8)\times 10^{19}$~eV. 

Let us now concentrate our attention on UHECR from superheavy dark matter
(SHDM) \cite{BKV97} and topological defects (TD) \cite{HiSch}.
The comparison of the UHECR spectrum obtained with the AGASA data, will 
provide us with the correct neutrino flux normalization.

Production of SHDM particles naturally occurs in a time-varying 
gravitational field of the expanding universe at the post-inflationary 
stage. The relic density of these particles is mainly determined
(at fixed reheating temperature and inflaton mass) by their
mass $M_X$. The range of practical interest is $(3 - 10)\times
10^{13}$ GeV, at larger masses the SHDM is a subdominant component of
the DM. SHDM is accumulated in the Galactic halo with the overdensity 
$\delta= \frac{\bar{\rho}_X^{\rm halo}}{\rho_X^{\rm extr}}=
\frac{\bar{\rho}_{\rm DM}^{\rm halo}}{\Omega_{\rm CDM}\rho_{\rm cr}}$,
where $ \bar{\rho}_{\rm DM}^{\rm halo}\approx 0.3$ GeV/cm$^3$, 
$\rho_{\rm cr}=1.88\times 10^{-29}h^2$ g/cm$^3$ and 
$\Omega_{\rm CDM}h^2=0.135$ \cite{WMAP}. With these numbers, 
$\delta \approx 2.1\times 10^5$. Because of this large local overdensity,
UHECRs from SHDM have no GZK cutoff.

Clumpiness of SHDM in the halo can provide the observed small-angle
clustering. The ratio $r_X=\Omega_X (t_0/\tau_X)$ of relic
abundance $\Omega_X$ and lifetime $\tau_X$ of the $X$ particle is
fixed by the observed UHECR flux as $r_X\sim 10^{-11}$. 
In the most interesting case of gravitational production of $X$
particles, their present abundance is determined by their
mass $M_X$ and the reheating temperature $T_R$.
Choosing a specific particle physics model one can fix also the life-time 
of the $X$ particle. There exist many models in which SH particles can
be quasi-stable with lifetime $\tau_X \gg 10^{10}$~yr.
The measurement of the UHECR flux, and thereby of $r_X$, selects from
the three-dimensional parameter space $(M_X, T_R, \tau_X)$ a
two-dimensional subspace compatible with the SHDM hypothesis.

\begin{figure}[ht]
\begin{center}
\begin{tabular}{ll}
\includegraphics[width=0.5\textwidth]{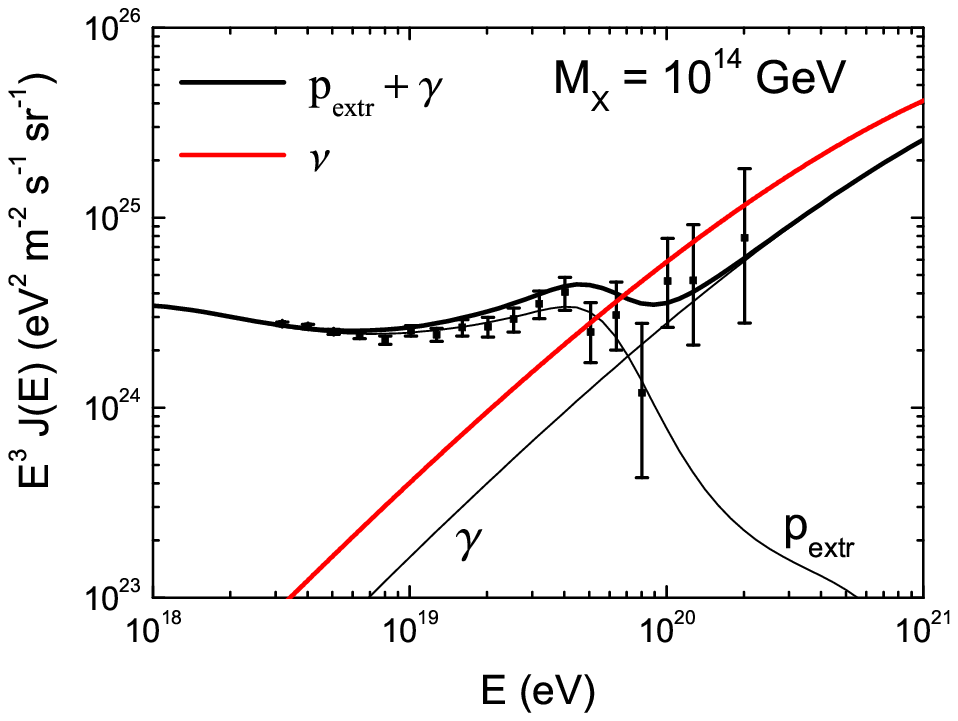}
&
\includegraphics[width=0.5\textwidth]{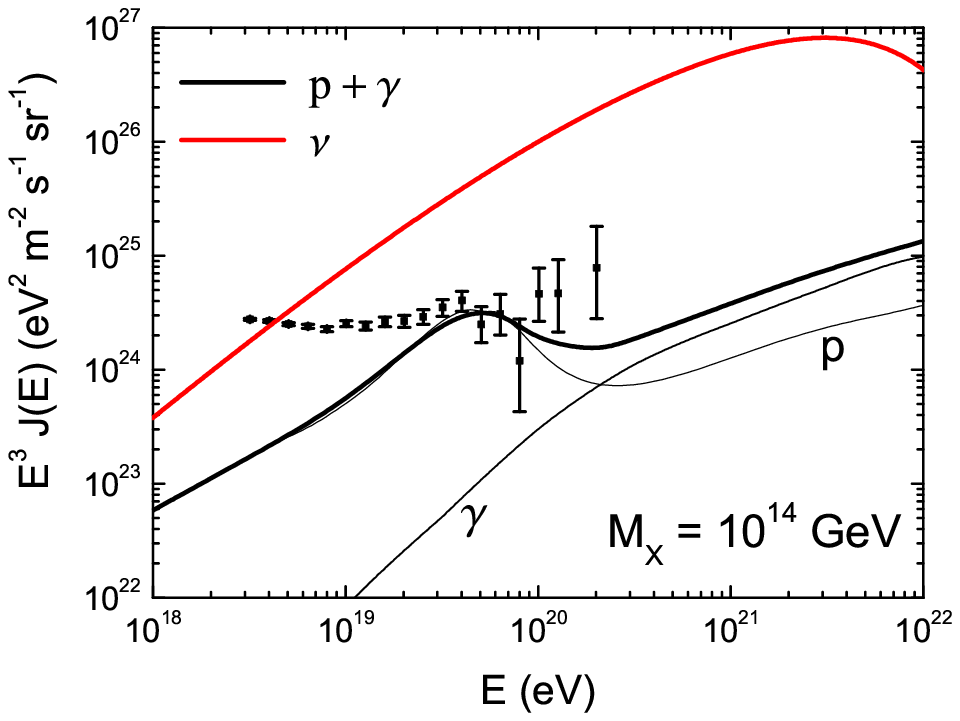}
\\
\end{tabular}
\caption{[Right Panel] Comparison of SHDM prediction with the AGASA data. 
The calculated spectrum of SHDM photons is shown by the label $\gamma$ and 
by the label $p_{extr}$ the spectrum of extragalactic protons (uniformly 
distributed astrophysical sources). The sum of these two spectra is shown 
by the thick black curve. The red thick line is the SHDM neutrino flux.
[Left Panel] Diffuse spectra from necklaces. The red thick curve shows 
neutrino flux, the black thick curve is the sum of protons and photons fluxes 
produced by necklaces (labeled thin black lines).}
\label{fig}
\end{center}
\end{figure}

In Figure \ref{fig} (left panel) we have performed a fit to the AGASA data 
using the photon flux from the SHDM model and the proton flux from uniformly 
distributed astrophysical sources. For the latter we have used the 
non-evolutionary model of \cite{dip}. The photon flux is normalized to 
provide the best fit to the AGASA data at $E\geq 4\times 10^{19}$~eV. 

One can see  from the fit of Figure \ref{fig} (left panel), that the SHDM
model can explain only the excess of AGASA events at 
$E \ge 1\times 10^{20}$~eV: depending on the SHDM spectrum
normalization and the details of the calculations for the extragalactic
protons, the flux from SHDM decays becomes dominant only  
above $(6-8)\times 10^{19}$~eV.

Topological Defects (for a review see \cite{Xpart} and reference therein)
can naturally produce UHE particles. 
The following TD have been discussed as potential sources of UHE
particles: superconducting strings, ordinary strings,
monopolonium (bound monopole-antimonopole pair), monopolonia
(monopole-antimonopole pairs connected by a string), networks of 
monopoles connected by strings, vortons and necklaces (see Ref.~\cite{Xpart}
for a review and references).
Monopolonia and vortons are clustering in the Galactic halo and their 
observational signatures for UHECR are identical to SHDM. However the friction 
of monopolonia in cosmic plasma results in monopolonium lifetime much 
shorter than the age of the universe. 
Of all other TD which are not clustering in the Galactic halo, the most 
favorable for UHECR are {\em necklaces}. 

Necklaces are hybrid TD produced in the symmetry breaking pattern
$G \rightarrow H \times U(1) \rightarrow H \times Z_2$. At the first
symmetry breaking monopoles are produced, at the second one each
(anti-) monopole get attached to two strings. This system resembles
ordinary cosmic strings with monopoles playing the role of
beads. Necklaces exist as the long strings and loops.
The symmetry breaking scales of the two phase transitions, $\eta_m$
and $\eta_s$, are the main parameters of the necklaces. They determine
the monopole mass, $m \sim 4\pi \eta_m/e$, and the mass of the string per
unit length $\mu \sim 2\pi \eta_s^2$. The evolution of necklaces is
governed by the ratio $r\sim m/\mu d$, where $d$ is the average
separation of a monopole and antimonopole along the string. As it is
argued in Ref.~\cite{neckl}, necklaces evolve towards configuration with
$r\gg 1$. Monopoles and antimonopoles trapped in the necklaces inevitably 
annihilate in the end, producing heavy Higgs and gauge bosons ($X$ particles) 
and then hadrons. The rate of $X$ particles production in the universe can be
estimated as \cite{neckl} $\dot{n}_X \sim \frac{r^2 \mu}{t^3 M_X}$,
where $t$ is the cosmological time.

The photons and electrons from pion decays initiate e-m cascades and
the cascade energy density can be calculated as
$\omega_{\rm cas}=\frac{1}{2}f_{\pi}r^2\mu \int_0^{t_0} \frac{dt}{t^3}
\frac{1}{(1+z)^4}=\frac{3}{4}f_{\pi}r^2\frac{\mu}{t_0^2}$,
where $z$ is the redshift and $f_{\pi}\sim 1$ is the fraction of the 
total energy release transferred to the cascade. 
The parameters of the necklace model for UHECR are restricted by the 
EGRET observations~\cite{EGRET} of the diffuse gamma-ray flux.  
This flux is produced by UHE electrons and photons from
necklaces due to e-m cascades initiated in collisions with CMB photons. In the
range of the EGRET observations, $10^2 - 10^5$~MeV, the predicted spectrum
is $\propto E^{-\alpha}$ with $\alpha=2$ \cite{cascade}. The EGRET
observations determined the spectral index as $\alpha=2.10\pm 0.03$
and the energy density of radiation as $\omega_{\rm obs} \approx
4\times 10^{-6}$~eV/cm$^3$. The cascade limit consists in the 
bound $\omega_{\rm cas}\leq \omega_{\rm obs}$.
According to the recent calculations, the Galactic contribution of gamma 
rays to the EGRET observations is larger than estimated earlier, and the 
extragalactic gamma-ray spectrum is not described by a power-law with 
$\alpha=2.1$. In this case, the limit on the cascade radiation with $\alpha=2$ 
is more restrictive and is given by 
$\omega_{\rm cas} \leq 2\times 10^{-6} {\rm eV/cm}^3$; we shall use this limit 
in further estimates. Using $\omega_{\rm cas}$ with $f_{\pi}= 1$ and 
$t_0=13.7$~Gyr \cite{WMAP} we obtain from the limit on the cascade radiation 
$r^2\mu \leq 8.9\times 10^{27}$~GeV$^2$.

The important and unique feature of necklaces is their small separation $D$, 
which ensures an high density. The distance $D$ is given by 
$D \sim r^{-1/2}t_0$ \cite{neckl}; since $r^2\mu$ is limited by e-m cascade 
radiation we can obtain a lower limit on the separation between necklaces as
$D \sim \left (\frac{3f_{\pi}\mu}{4t_0^2\omega_{\rm cas}} \right )^{1/4}t_0
> 10(\mu/10^6~{\rm GeV}^2)^{1/4}~{\rm kpc}$,
this small distance is a unique property of necklaces allowing the
unabsorbed arrival of particles with the highest energies.
The fluxes of UHECR from necklaces are shown in Figure \ref{fig} (right panel).
We used in the calculations $r^2\mu = 4.7\times 10^{27}$~GeV$^2$ which 
corresponds to $\omega_{\rm cas}= 1.1\times 10^{-6}$~ eV/cm$^3$, i.e. 
twice less than allowed by the bound on $\omega_{\rm cas}$. The mass of the $X$
particles produced by monopole-antimonopole annihilations is taken as  
$M_X= 1\times 10^{14}$~GeV. From  Figure \ref{fig} (right panel) one can see 
that the necklaces model for UHECR can explain only the highest energy part 
of the spectrum, with the AGASA excess somewhat above the prediction. Thus 
UHE particles from necklaces can serve only as an additional component in the 
observed UHECR flux. This result has a particular impact on the possible UHE 
neutrino detection. In fact, the necklaces model is only under constrained by 
the available UHECR data, in this context only a clear UHE neutrino 
observation with a typical spectrum as in figure \ref{fig} (right panel) 
can confirm (or falsify) the model.
  
\section*{Acknowledgments}
{\small 
I am grateful to V. Berezinsky and M. Kachelrie{\ss} with whom the 
present work was developed. 
}

\setcounter{section}{0}
\setcounter{subsection}{0}
\setcounter{figure}{0}
\setcounter{table}{0}
\newpage

\begin{thebibliography}{00}
{\small

\bibitem{GZK}
K. Greisen, Phys. Rev. Lett. {\bf 16}, 748 (1966);
G.T. Zatsepin and V.A. Kuzmin, JETP Lett.  {\bf 4}, 78 (1966)
[Pisma Zh. Eksp. Teor. Fiz. {\bf 4}, 114 (1966)]. 

\bibitem{expCR}
M. Takeda {\it et al.} [AGASA collaboration], astro-ph/0209422. 
N. Hayashida {\it et al.} [AGASA collaboration], 
Phys. Rev. Lett. {\bf 73}, 3491 (1994).
K. Shinozaki {\it et al.} [AGASA collaboration],
Astrophys. J. {\bf 571}, L 117 (2002).
T. Abu-Zayyad {\it et al.} [HiRes collaboration], astro-ph/0208243.
D.J. Bird {\it et al.} [Fly's Eye collaboration], 
Ap.J. {\bf 424}, 491 (1994).
J. Bl\"umer {\it et al.} [Auger Collaboration], J. Phys. {\bf G29} 867 (2003).

\bibitem{TT}
P.G. Tinyakov and I.I. Tkachev, JETP Lett., {\bf 74}, 445 (2001);
astro-ph/0301336 (and references therein).

\bibitem{Hillas}
M.A. Hillas J. Phys. {\bf G31} 95 (2005)

\bibitem{BereZats}
V.S. Berezinsky and G.T. Zatsepin, Phys. Lett. {\bf B28} 6 (1969).

\bibitem{nu}
V.S. Berezinsky and G.T. Zatsepin, Phys. Lett. {\bf B28}, 423 (1969);
Z.~Fodor, S.D.~Katz, A.~Ringwald, and H.~Tu, 
Phys. Lett. B {\bf 561}, 191 (2003),
P. Jain, D.W. McKay, S. Panda and J.P. Ralston, 
Phys. Lett. {\bf B484}, 267 (2000); but see also
M.~Kachelrie{\ss}\ and M.~Pl\"umacher,
Phys.\ Rev.\ D {\bf 62}, 103006 (2000).

\bibitem{gluino}
D.J.H. Chung, G.R. Farrar and E.W. Kolb, 
Phys. Rev. D {\bf 57}, 4606 (1998);
V.S. Berezinsky and M. Kachelrie{\ss}, 
Phys. Lett. {\bf B422}, 163 (1998);
I.F.~Albuquerque, G.R.~Farrar and E.W.~Kolb,
Phys.\ Rev.\ D {\bf 59}, 015021 (1999);
V.S. Berezinsky, M. Kachelrie{\ss} and S. Ostapchenko, Phys. Rev. D {\bf 65} 
083004 (2002);
M.~Kachelrie\ss, D.~V.~Semikoz and M.~A.~T\'ortola,
Phys.\ Rev.\ D {\bf 68}, 043005 (2003).

\bibitem{Z}
D. Fargion, B. Mele and A. Salis, Ap. J. {\bf 517}, 725 (1999);
T.J. Weiler, Astrop. Phys. {\bf 11}, 303 (1999);
Z. Fodor, S.D. Katz and A. Ringwald, astro-ph/0203198.

\bibitem{Lorentz}
D.A. Kirzhnits and V.A. Checin, Sov. J. Nucl. Phys. {\bf 15}, 585 (1971);
S. Coleman and S.L. Glashow, Phys. Rev. {\bf D59}, 116008 (1999);
R. Aloisio, P. Blasi, P.L. Ghia and A.F. Grillo, 
Phys. Rev. {\bf D62}, 053010 (2000).

\bibitem{TD}
V.S. Berezinsky, Nucl. Phys. (Proc. Suppl) {\bf B87}, 387 (2000).

\bibitem{SHDM}
V.A. Kuzmin and I.I. Tkachev, Phys. Rep. {\bf 320}, 199 (1999).

\bibitem{GHT}
T.K. Gaisser, F. Halzen and T. Stanev, Phys. Rept. {\bf 258} 173 (1995)
[Erratum {\bf 271} 355 (1995)]. F. Halzen and D. Hooper, Rept. Prog. Phys. 
{\bf 65} 1025 (2002).
 
\bibitem{ABK}
R. Aloisio, V. Berezinsky and M. Kachekrie{\ss}, Phys. Rev. {\bf D69} 094023
(2004).

\bibitem{BereKach}
V. Berezinsky and M. Kachelrie{\ss}, Phys.\ Rev. {\bf D63}, 034007 (2001).

\bibitem{DGLAP}
V.N. Gribov and L.N. Lipatov, Sov. J. Nucl. Phys. {\bf 15}, 438 (1972);
Yu. L. Dokshitzer, Sov. Phys. JETP {\bf 46}, 641 (1977).
G. Altarelli and G. Parisi, Nucl. Phys. {\bf B126}, 298 (1977).

\bibitem{previous}
N.A. Rubin, Thesis, Cavendish Laboratory, University of Cambridge (1999).
S. Sarkar and R. Toldr\`a, Nucl. Phys. {\bf B621}, 495 (2002).
Z. Fodor and S.D. Katz, Phys. Rev. Lett. {\bf 86}, 3224 (2001).
C. Barbot and M. Drees, Astropart. Phys. {\bf 20}, 5 (2003).

\bibitem{Kniehl}
B.~A.~Kniehl, G.~Kr\"amer and B.~P\"otter, Nucl. Phys. {\bf B 582}, 514 (2000).

\bibitem{BKV97}
V. Berezinsky, M. Kachelrie{\ss} and A. Vilenkin, 
Phys. Rev. Lett. {\bf 79}, 4302 (1997). 
V.A. Kuzmin and V.A. Rubakov, Phys. At. Nucl. {\bf 61}, 1028 (1998). 

\bibitem{HiSch}
C.T. Hill, D.N. Schramm and T.P. Walker, Phys. Rev. {\bf D36}, 1007 (1987).

\bibitem{WMAP}
D.N. Spergel {\it et al.} [WMAP collaboration], astro-ph/0302209.

\bibitem{dip}
R. Aloisio, V. Berezinsky, P. Blasi, A. Gazizov and S. Grigorieva,
astro-ph/0608219. 

\bibitem{Xpart}
V.A. Kuzmin and I.I. Tkachev, Phys. Rep. {\bf 320}, 199 (1999).

\bibitem{neckl}
V.S. Berezinsky and A. Vilenkin, Phys. Rev. Lett. {\bf 79}, 5202 (1997).

\bibitem{EGRET}
P. Sreekumar et al, Ap.J., {\bf 494}, 523, (1998).

\bibitem{cascade}
V.S.Berezinsky and A.Yu.Smirnov, Ap.Sp.Sci. {\bf 32}, 463, (1975).

}
\end{thebibliography}
\end{document}